\documentclass[prl,twocolumn,amsmath,amssymb,floatfix,reprint,footinbib,superscriptaddress,longbibliography,showkeys]{revtex4-1}
\usepackage{dcolumn}
\usepackage{graphicx}
\usepackage{mathrsfs}
\usepackage{mdwlist}
\usepackage{subfigure}
\usepackage{booktabs}
\usepackage{amsmath}
\usepackage{dsfont}
\usepackage{amstext}
\usepackage{amssymb}
\usepackage{amsbsy}
\usepackage{bbm}
\usepackage{amsthm}
\usepackage{graphicx}
\usepackage{textcomp}
\usepackage{multirow}
\usepackage{color}
\usepackage[colorlinks,citecolor=blue]{hyperref}
\definecolor{Dgreen}{RGB}{0, 100, 0}
\usepackage{url}
\usepackage[colorlinks]{hyperref}
\hypersetup{%
	plainpages=true,
	breaklinks=true,  
	hypertexnames=false,  
	pageanchor=true,
	colorlinks=true,
	linkcolor={blue},
	citecolor={red},
	urlcolor={blue},
	anchorcolor={black}
}

\begin{document}
	
	\title{Shortcuts to Adiabaticity for the Quantum Rabi Model: Efficient Generation of Giant Entangled cat States via Parametric Amplification}
	\author{Ye-Hong Chen}
	\affiliation{Theoretical Quantum Physics Laboratory, RIKEN Cluster for Pioneering Research, Wako-shi, Saitama 351-0198, Japan}
	
	\author{Wei Qin}
	\affiliation{Theoretical Quantum Physics Laboratory, RIKEN Cluster for Pioneering Research, Wako-shi, Saitama 351-0198, Japan}
	
	\author{Xin Wang}
	\affiliation{Theoretical Quantum Physics Laboratory, RIKEN Cluster for Pioneering Research, Wako-shi, Saitama 351-0198, Japan}
	\affiliation{Institute of Quantum Optics and Quantum Information, School of Science, Xi'an Jiaotong University, Xi'an 710049, China}
	
	\author{Adam Miranowicz}
	\affiliation{Theoretical Quantum Physics Laboratory, RIKEN Cluster for Pioneering Research, Wako-shi, Saitama 351-0198, Japan}
	\affiliation{Faculty of Physics, Adam Mickiewicz University, 61-614 Pozna\'{n}, Poland}
	
	\author{Franco Nori}
	\affiliation{Theoretical Quantum Physics Laboratory, RIKEN Cluster for Pioneering Research, Wako-shi, Saitama 351-0198, Japan}
	\affiliation{Department of Physics, University of Michigan, Ann Arbor, Michigan 48109-1040, USA}

	\date{\today}

	\begin{abstract}
		We propose a method for the fast generation of nonclassical ground states of the Rabi model in the ultrastrong and deep-strong coupling regimes via the shortcuts-to-adiabatic (STA) dynamics.
		The time-dependent quantum Rabi model is simulated by applying parametric amplification to the Jaynes-Cummings model. Using experimentally feasible parametric drive, this STA protocol can generate large-size Schr\"{o}dinger cat states, through a process that is $\sim 10$ times faster compared to adiabatic protocols.
		Such fast evolution increases the robustness of our protocol against
		dissipation.
		Our method enables one to freely design the parametric drive, so that the target state can be generated in the lab frame. A largely detuned light-matter coupling makes
		the protocol robust against imperfections of the operation times in experiments.
	\end{abstract}
	
	
	\maketitle

	\textit{Introduction.}---The quantum Rabi model \cite{Pr49324,Pr51652,Prl107100401} is the lowest-dimensional Hamiltonian describing the light-matter interaction beyond the rotating-wave approximation (RWA),
	\begin{align}\label{eq1-1}
	{H}_{\rm{R}}=\omega_{c} {a}^{\dag}{a}+\frac{\omega_{q}}{2}{\sigma}_{z}+{\sigma}_{x}(g{a}^{\dag}+g^{*}{a}).\ \ \ \ \ (\hbar=1)
	\end{align}
	Here, $\omega_{c}$ $(\omega_{q})$ is the frequency of the cavity (qubit), $g$ is the light-matter coupling strength,
	${a}^{\dag}$ (${a}$) is the creation (annihilation) operator of the cavity field,
	and ${\sigma}_{x}$ and
	${\sigma}_{z}$
	are Pauli operators of the qubit.
	This model was first introduced 90 years ago and it has been used to describe the dynamics of
	a wide variety of physical setups \cite{kwek2013strong}, ranging from quantum optics
	to condensed matter physics.
	{The popular models of Dicke \cite{Pr9399}, Hopfield \cite{Pr1121555},
		and Tavis-Cummings \cite{Pr170379} are just multiqubit generalizations
		of the Rabi model, while the Jaynes-Cummings (JC) model \cite{Pi5189} is its
		simplified version \cite{Nr119}.}
	Generally, the Rabi model can be divided into different coupling regimes \cite{Nr119,Rmp91025005,Pra96013849}, according to the normalized coupling strength $\eta=g/\omega_{c}$.
	When focusing on the ultrastrong ($|\eta|\simeq0.1\sim 1$) and deep-strong ($|\eta|\gtrsim 1$) regimes,
	the counterrotating terms in ${H}_{\rm{R}}$ cannot be neglected. This leads to
	areas of unexplored physics and gives
	rise to many fascinating quantum phenomena, such as the asymmetry
	of vacuum Rabi splitting \cite{Njp13073002}, nonclassical photon statistics
	\cite{Pra87013826,Pra81042311}, and superradiance transition
	\cite{Prl109193602,Prl110163601,Pra96023863,Pra98063815}.
	
	For instance, the ground state of the Rabi model
	is a squeezed-vacuum state and involves virtual cavity photons \cite{Pra81042311,Pra70022303,Pra96043834}.
	Specifically, when $\omega_{q}\ll g$, the ground state of
	the Rabi model is
	\begin{align}\label{eq1-3}
	|G\rangle=&\frac{1}{2}\left(\mathcal{N}_{+}|g\rangle|\text{cat}_{+}\rangle
	-\mathcal{N}_{-}|e\rangle|\text{cat}_{-}\rangle\right),
	\end{align}
	which is an entangled Schr\"{o}dinger cat state.
	Here, $\mathcal{N}_{\pm}=\sqrt{2\left[1\pm \exp{\left(-2|\eta|^2\right)}\right]}$ determine the probability amplitudes of the even ($+$) and odd ($-$) cat states
	$|\text{cat}_{\pm}\rangle=(|\eta\rangle\pm|-\eta\rangle)/\mathcal{N}_{\pm}$, respectively. The states $|\pm\eta\rangle$ are coherent states.
	The state $|g\rangle$ $(|e\rangle)$ is the ground (excited) state
	of the qubit.
	By imposing the system to be in this ground state, one can generate
	the maximally entangled cat state (MECS) when $\mathcal{N}_{+}\simeq \mathcal{N}_{-}$.
	{The generation of the MECSs is significant not only for the demonstration of the fundamentals of quantum
		physics, but also has wide applications in modern quantum technologies, such as quantum
		information processing \cite{Npton13110,Prl119030502,Njp16045014,Prl116140502,Pra101043841} and quantum metrology \cite{Prl107083601}. For instance, giant cat qubits are very robust against photon dephasing, so that
		they can be very promising for fault-tolerant
		quantum computation \cite{Prl119030502,Njp16045014,Prl116140502}. }
	
	
	To generate the MECS,
	the system needs to enter the deep-strong coupling (DSC) regime of $|\eta|\gtrsim \sqrt{2}$,
	which is, however, still difficult to achieve in experiments \cite{Np6772,Prl105237001,npjQI346,Np1339,Np1344,Sr626720,Prb93214501,Pra96012325,Pra95053824,Prl120183601}.
	Researchers are encouraged to use simulation protocols \cite{Prx2021007,Sr407482,Pra102033716,Nc81715,Prl120093601,Prl120093602,Prl114093602,Pra100012339,Sci3641163,Q4271,arXiv200914342,Nap94853} based on the  JC model \cite{Jmo401195,Scully1997,Agarwal2009} to study
	exotic phenomena in the DSC regime.
	For instance, using linear \cite{Prx2021007} or
	nonlinear drives \cite{Prl120093601,Prl120093602}, one can
	modify the sideband of a cavity-qubit coupled system, so as to enhance
	the effective light-matter coupling to enter the
	DSC regime.
	This opens the possibility to adiabatically control the effective coupling strength
	based on, e.g., a time-dependent parametric drive,
	to prepare the target state $|G\rangle$ in the squeezed-light frame \cite{Prl120093602}. However, the adiabatic control requires a very small changing rate
	in the control parameters, usually leading to a long-time evolution.
	Such a long-time evolution inevitably increases the effect
	of dissipation, resulting in a low-fidelity target state.
	In addition, how to turn off the parametric drive without affecting the prepared entangled state is still an open problem.

	In this Letter,  we propose to use shortcuts-to-adiabatic (STA) methods
	\cite{Rmp91045001,Aamo62117,Prl104063002,Jpca1079937,Jpa42365303,Prl105123003,Prl111100502,Prl116230503,Pra93052109,Pra95042345,Sr622202,Pra99032323}, e.g.,
	counterdiabatic (CD) driving,
	to rapidly generate the target state $|G\rangle$.
	The STA methods are a series of protocols mimicking adiabatic
	dynamics beyond the adiabatic limit and have been widely applied for quantum state engineering \cite{Prl851626,Prl109115703,Prl124180401,Pra90060301,Pra95062319,Njp20015010,Prl109100403,njpQI31,Prl124150603,Nc712479,Np13330,Sa5eaau5999,Prl122090502,Prl114177206,Nc711338,arXiv200903539}.
	Specifically, the CD driving \cite{Jpca1079937,Jpa42365303} enables controlling a quantum system, such that the system can accurately evolve along an adiabatic path (e.g., an instantaneous eigenstate of the reference Hamiltonian)
	beyond the adiabatic limit, where nonadiabatic excitations can be
	precisely compensated by, e.g., adding an auxiliary driving term to a reference Hamiltonian \cite{Pr6971}.
	Using the STA method allows us to significantly shorten the evolution
	time as compared to the
	adiabatic protocol.
	Thus, we can suppress the effect of dissipation and significantly
	improve the fidelity of a given state.
	The parametric drive can be smoothly turned off in our STA protocol,
	because the amplitudes of the parametric drive are continuously turnable.
	{Additionally, the discussed model is
		generic, so our proposal can be realized in many
		physical systems, in particular, circuit quantum electrodynamics (QED) systems \cite{Prl120093601,Prl120093602,Nc711338} or ion traps \cite{Sci3641163,arXiv200914342}.}

	\textit{Adiabatic limit.}---Assuming
	$\omega_{q}\ll g$ and $H_{\rm{R}}\equiv H_{\rm{R}}(t)$ [with a controllable parameter $\eta\equiv\eta(t)$] to be time dependent,
	the Rabi Hamiltonian in Eq.~(\ref{eq1-1}) can be diagonalized by
	the unitary operator \cite{Nr119,Rmp91025005}
	\begin{align}\label{eq1-4}
	{U}(t)=|{+_{x}}\rangle\langle{+_{x}}|{D}[-\eta(t)]+|{-_{x}}\rangle\langle{-_{x}}|{D}[\eta(t)],
	\end{align}
	where $|\pm_{x}\rangle$ are the eigenstates of $\sigma_{x}$ and ${D}[\eta(t)]=\exp[\eta(t){a}^{\dag}-\eta^{*}(t){a}]$
	is the displacement operator.
	{To avoid the nonadiabatic transitions between
		the instantaneous eigenstates $\{|E_{m}(t)\rangle\}$ [eigenvalues $\xi_{m}(t)$]
		of $H_{\rm{R}}(t)$, the system needs to satisfy the adiabatic condition
		$|\langle E_{m}(t)|\dot{E}_{n\neq m}(t)\rangle|\ll |\xi_{m}(t)-\xi_{n}(t)|$.}

	\textit{CD-driving Hamiltonian.}---According to Eq.~(\ref{eq1-4}) and Berry's transitionless algorithm \cite{Jpa42365303},
	the CD-driving Hamiltonian for the reference Hamiltonian $H_{\rm{R}}(t)$ is
	\begin{align}\label{eq1-6}
	{H}_{\rm{CD}}(t)=i\dot{U}(t){U}^{\dag}(t)
	=i{\sigma}_{x}[\dot{\eta}^{*}(t){a}-\dot{\eta}(t){a}^{\dag}].
	\end{align}
	The desired STA process can be realized by adding the CD-driving Hamiltonian
	$H_{\rm{CD}}(t)$ into the reference Hamiltonian $H_{\rm{R}}(t)$
	to construct a feasible total Hamiltonian: i.e., $H_{\rm{tot}}(t)=H_{\rm{R}}(t)+H_{\rm{CD}}(t)$ \cite{Prl105123003}.
	In this case, we can predict an ideal evolution along the
	instantaneous eigenstate $|E_{m}(t)\rangle$, as ${H}_{\rm{tot}}(t)$ ideally satisfies the Schr\"{o}dinger equation
	$i|\dot{E}_{m}(t)\rangle=[\xi_{m}(t)+H_{\rm{CD}}(t)]|E_{m}(t)\rangle$ \cite{Pra95062319}.
	Thus, assuming the initial state to be $|E_{0}(0)\rangle=|g\rangle|0\rangle$,
	we obtain the target state $|E_{0}(t_{f})\rangle=|G\rangle$ at the final time $t_{f}$.
	However, realizing a time-dependent Rabi model in the DSC
	regime is still a major challenge in experiments.
	In the following, we illustrate how to simulate $H_{\rm{tot}}(t)$
	based on a parametrically driven JC model, so as to
	realize the STA protocol and generate the state $|G\rangle$.

	\begin{figure}
		\centering
		\scalebox{0.33}{\includegraphics{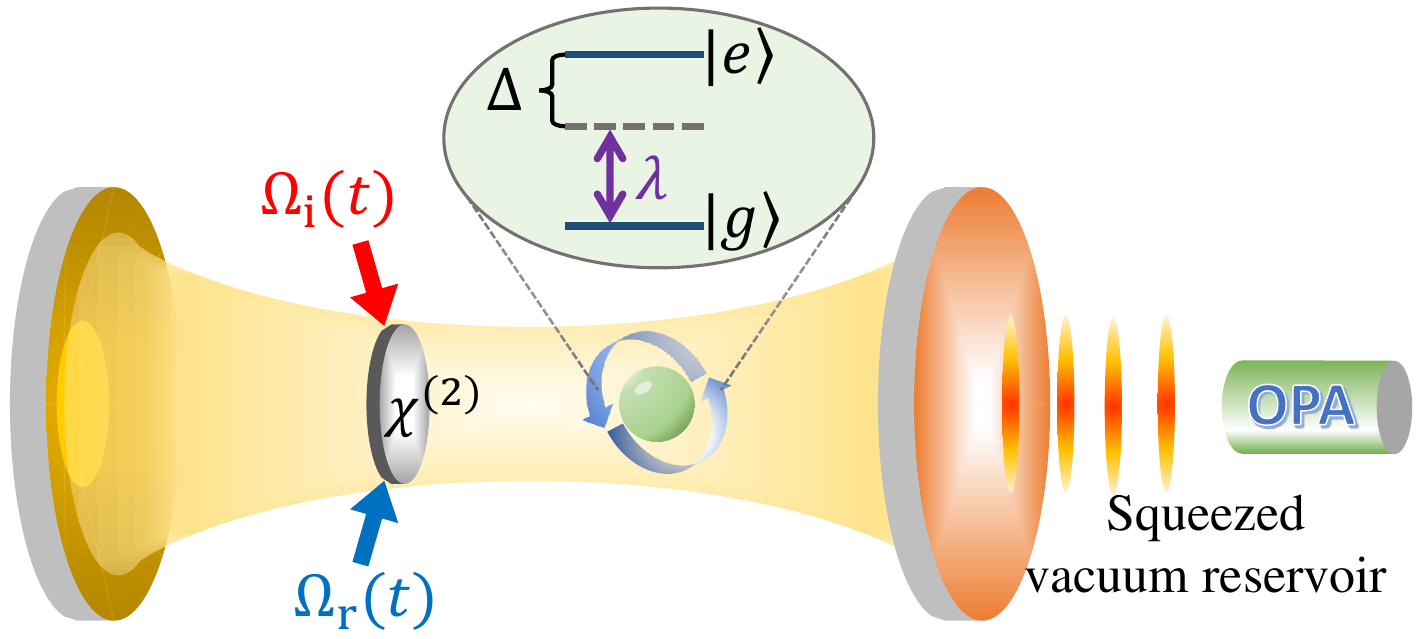}}
		\caption{Schematic illustration of a cavity QED
			system containing a single-mode cavity, a qubit, a $\chi^{(2)}$-nonlinear medium, and an optical parametric amplifier (OPA). The qubit ($|g\rangle$=ground state and $|e\rangle$=excited state) coupled to the cavity with coupling strength $\lambda$ and large detuning $\Delta$.
			The nonlinear drive $\Omega_{\rm{r}}(t)$ induces a time-dependent squeezed-cavity mode. The other nonlinear drive $\Omega_{\rm{i}}(t)$ [$\pi/2$-dephased from $\Omega_{\rm{r}}(t)$] is applied to counteract the nonadiabatic transition induced by mapping the system dynamics into the time-dependent squeezed-light frame. The OPA is used to generate a squeezed-vacuum reservoir, which couples to the cavity mode to minimize the influence of the squeezing-induced noise.
		}
		\label{fig00}
	\end{figure}

	\textit{Model and effective Hamiltonian.}---As shown in Fig.~\ref{fig00}, our STA proposal is realized
	in the JC model.
	The
	cavity is subjected to two time-dependent (two-photon) drives,
	with the same frequency $\omega_{p}$, but with different real amplitudes, $\Omega_{\rm{r}}(t)$ and $\Omega_{\rm{i}}(t)$.
	The drive $\Omega_{\rm{i}}(t)$ is $\pi/2$-dephased from $\Omega_{\rm{r}}(t)$.
	The Hamiltonian in a frame rotating at $\omega_{p}/2$ reads
	\begin{align}\label{eq1-7}
	{H}_{0}(t)=\Delta{a}^{\dag}{a}
	-\left[\frac{\Omega_{\rm{r}}(t)+i\Omega_{\rm{i}}(t)}{2}{a}^{2}-\lambda{a^{\dag}}\sigma+\text{h.c.}\right],
	\end{align}
	{where $\Delta=\omega_{c}-\omega_{p}/2$, $\sigma=|g\rangle\langle e|$, $\lambda\ll\omega_{\rm{c}/\rm{q}}$ is the qubit-cavity coupling strength,
		and we have assumed $\omega_{q}=\omega_{p}$.}
	By performing the unitary transformation ${S}(t)=\exp{[r(t)({a}^{\dag 2}-{a}^{2})/2]}$, with $r(t)$ satisfying $\tanh\left[2r(t)\right]=\Omega_{\rm{r}}(t)/\Delta$, we obtain the effective Hamiltonian
	\begin{align}\label{eq1-8}
	{H}_{S}(t)&\approx\Delta\text{sech}{[2r(t)]}a^{\dag}a+\lambda{e^{r(t)}}{\sigma_{x}}( a^{\dag}+a)/2,
	\end{align}
	where we have neglected the undesired terms
	by assuming $\Omega_{\rm{i}}(t)=\dot{r}(t)$ and $\lambda\ll\Delta$.
	The condition $\Omega_{\rm{i}}(t)=\dot{r}(t)$ has been applied according to the transitionless
	algorithm to counteract
	the nonadiabatic transition caused by the time-dependent unitary transformation $S(t)$ (see the Supplemental Material \cite{SM} for details).
	The effective normalized coupling
	strength of ${H}_{S}(t)$ is
	\begin{align}\label{eq1-9}
	\tilde{\eta}(t)=\frac{{\lambda}}{4\Delta}
	\left\{\exp{[3{r}(t)]}+\exp{[-{r}(t)]}\right\}.
	\end{align}
	{To show the advantages of our STA protocol, as compared to the adiabatic scheme \cite{Prl120093602},
		in the following discussion}
	we denote $\tilde{*}$ and $*$ ($*=\eta,\lambda,r,\ldots$) to represent all the parameters in the adiabatic and STA processes, respectively. Here, $\tilde{*}$ and $*$ have the same
	physical meaning.
	
\begin{figure}
		\centering
		\scalebox{0.32}{\includegraphics{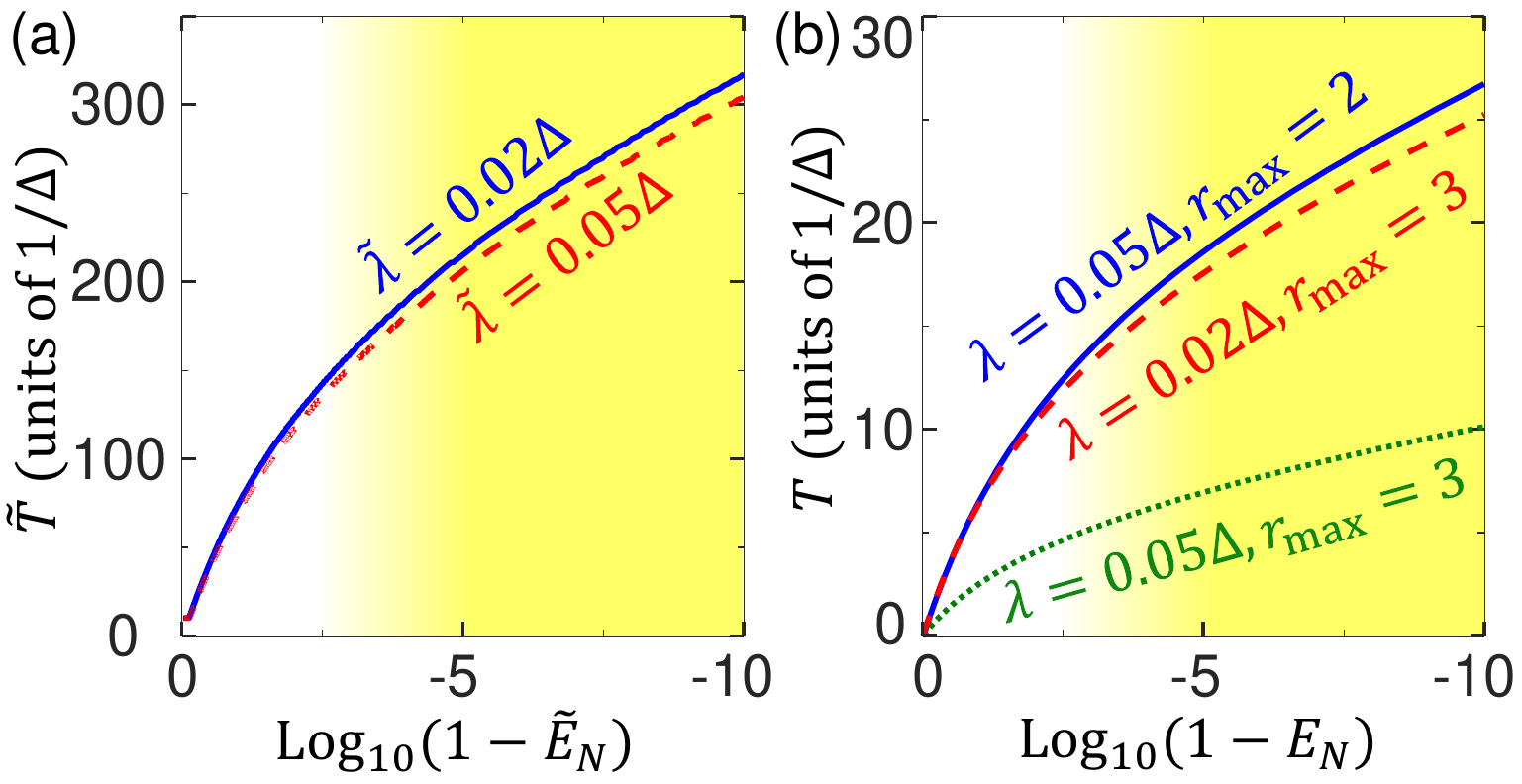}}
		\caption{Total evolution time required for (a) the adiabatic process and (b) the STA process,
			to achieve the MECS versus the desired entanglement cost [characterized by the
			logarithmic negativity $\tilde{E}_{N}$ ($E_{N}$)]. For (a), the squeezing parameter is
			$\tilde{r}(t)=\tilde{r}_{\rm{max}}/\{1+\exp{[\tilde{f}(t)]}\}$, where $\tilde{r}_{\rm{max}}$ is the
			peak value of $\tilde{r}(t)$ and $\tilde{f}(t)=\tilde{f}_{0}(1/2-t/\tilde{T})$, with $\tilde{f}_{0}=10$.
			For (b), the parameters are $\eta(0)=(1+i)/100$ and ${r}(t)={r}_{\rm{max}}/\{1+\exp{[{f}(t)]}\}$, with $f(t)=f_{0}\cos{(2\pi t/T)}$ and
			$f_{0}=10$, resulting in
			$r(0)=r(t_{f})\simeq0$ and $\dot{r}(0)=\dot{r}(t_{f})\simeq 0$. The light-matter coupling ($\tilde{\lambda},~\lambda\ll\Delta$) are chosen to satisfy
			the condition to neglect the undesired terms to obtain the effective Hamiltonian in Eq.~(\ref{eq1-8}).
			The comparison between the panels shows that, the time required in the STA process to achieve the target state is \textit{$\sim10$ times shorter} than that required in the adiabatic process.
			The yellow-shaded area in each panel shows $(\tilde{E}_{N},\ E_{N}\gtrsim 99\%)$, indicating that the target state in this area is maximally entangled.
		}
		\label{fig0}
	\end{figure}

	\textit{Adiabatic protocol.}---{When $|\dot{\tilde{\eta}}(t)|\ll\Delta\text{sech}\left[2\tilde{r}(t)\right]$, one can achieve the adiabatic evolution along the ground eigenstate of $H_{S}(t)$ \cite{Prl120093602}.}
	The adiabatic condition requires $\dot{\tilde{r}}(t)/\Delta\rightarrow 0$, thus leading to slow evolution.
	Figure \ref{fig0}(a) shows the relationship between the total evolution time $\tilde{T}$ and
	the logarithmic negativity $\tilde{E}_{N}=\log_{2}||\rho^{\Gamma_{\rm{q}}}||_{1}$ \cite{Rmp81865} of the adiabatic process.
	Here, ${\Gamma_{\rm{q}}}$ denotes the partial
	transpose with respect to the qubit, and $||\cdot||_{1}$ the trace norm.
	The evolution time $\tilde{T}$ significantly increases when
	the desired entanglement cost grows. To achieve the MECS with $\tilde{E}_{N}\gtrsim 99.99\%$, one needs $\tilde{T}\gtrsim 200/\Delta$ via the adiabatic process.
	
	According to Eq.~(\ref{eq1-9}), a fixed
	final squeezing parameter $\tilde{r}(t_{f})=\tilde{r}_{\rm{max}}$ is needed
	to obtain the target state $|G\rangle$.
	As a result, the MECS only can
	be prepared in the squeezed-light frame rather than the
	lab frame, i.e., the final state is $S(t_{f})|G\rangle$.
	To obtain a MECS in the lab frame,
	one needs to turn off the parametric drive immediately when $t>t_{f}$.
	However, rapidly decreasing the squeezing parameter $r(t)$ induces an undesired nonadiabatic transition,
	which pumps many photons into the cavity in a very short time \cite{SM}.
	Then, the final state might be unpredictable.


	\textit{STA protocol.}---{We assume $H_{\rm{tot}}(t)=H_{S}(t)$, resulting in
		$\Delta\text{sech}[2r(t)]:\Rightarrow \omega_{c}$ and $\lambda \exp{[r(t)]}:\Rightarrow 2[g-i\dot{\eta}(t)]$, where $\eta(t)=g/\omega_{c}$. Thus, we
		obtain the equations of motion for the coherent state amplitude $\eta(t)$:
		\begin{align}\label{eq1-10}
		\text{Re}[\dot{\eta}(t)]=&\Delta\text{Im}[\eta(t)]\text{sech}2r(t),\cr
		\text{Im}[\dot{\eta}(t)]=&\frac{\lambda}{2} \exp[{r(t)}]-\Delta\text{Re}[\eta(t)]\text{sech}2r(t),
		\end{align}
		where $\text{Re}[*]$ ($\text{Im}[*]$) denotes the real (imaginary) part of the parameter ``$*$.'' Note that $\eta(t)=g/\omega_{c}$ is different from the definition of $\tilde{\eta}(t)$ in Eq.~(\ref{eq1-9}), thus the Hamiltonian $H_{S}(t)$ can drive the system to evolve along the ground eigenstate $|E_{0}(t)\rangle$ of the Hamiltonian $H_{\rm{R}}(t)$.}
	According to Eq.~(\ref{eq1-10}), $\eta(t)$ relies on the time integration of the
	squeezing parameter $r(t)$. This allows to rapidly achieve
	a large value of $\eta(t_{f})$ without any restrictions on the final squeezing parameter $r(t_{f})$.
	Thus, the STA process can achieve the target state $|G\rangle$ in the lab frame, i.e., $r(t_{f})=0$.

	In Fig.~\ref{fig0}(b), we display the total evolution time $T$ required for the STA
	process to obtain the target state versus the logarithmic negativity $E_{N}$.
	We find that $T$ is significantly shortened when we
	increase the coupling strength $\lambda$ and the peak squeezing parameter $r_{\rm{max}}$.
	For an experimentally feasible gain of $10\log_{10}[\exp{(2 r_{\rm{max}})}]\sim20$~dB \cite{Nat541191,Nat49962,Prl121173601}
	(corresponding to $r_{\rm{max}}\sim 2.3$), the evolution time to
	achieve the MECS with $E_{N}\gtrsim 99.99\%$ via
	the STA process is $T\sim 20/\Delta$, which is $\sim10$ times shorter than that via the adiabatic
	process.

	\begin{figure}
		\centering
		\scalebox{0.36}{\includegraphics{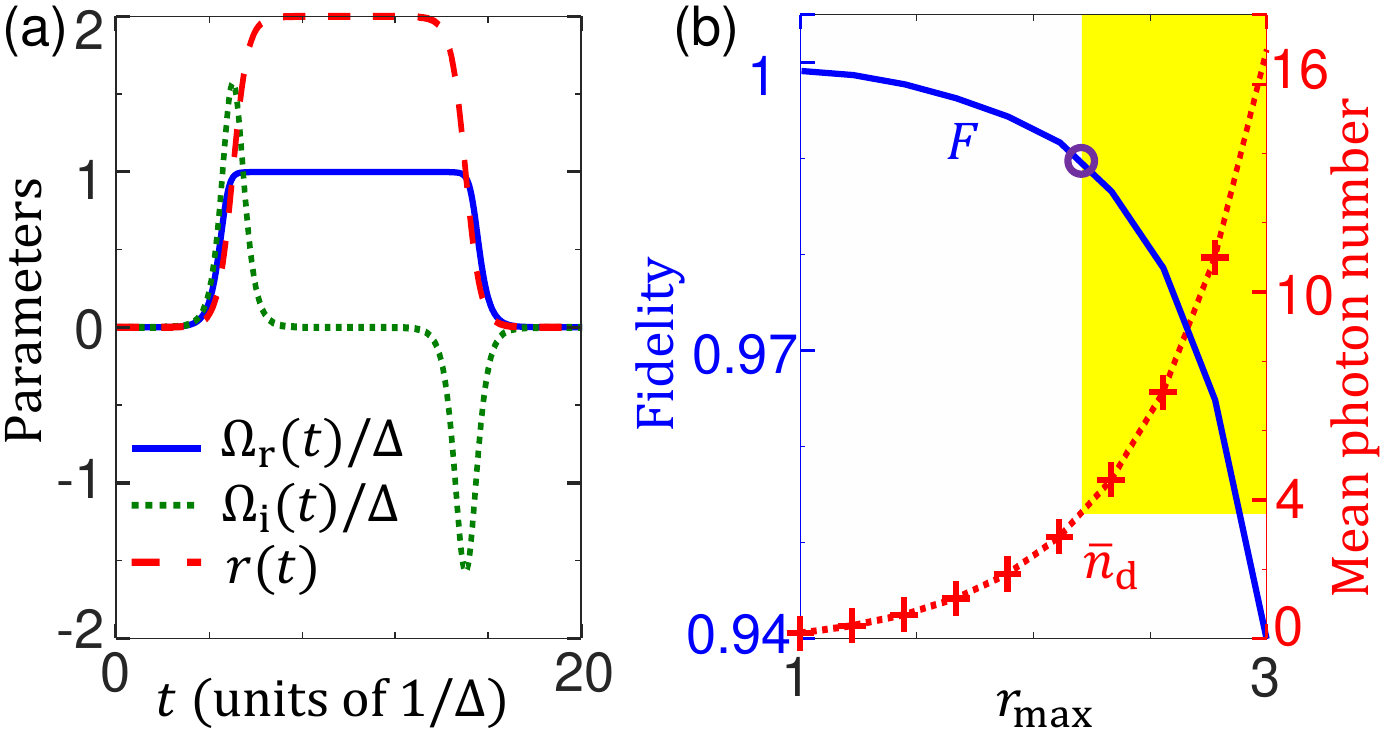}}
		\caption{{STA protocol when $\eta(0)=(1+i)/100$, $\lambda=0.045\Delta$, and $T=20/\Delta$: (a) Finite-duration drives $\Omega_{\rm{r}}(t)$, $\Omega_{\rm{i}}(t)$ and
				squeezing parameter $r(t)$ when the peak squeezing parameter $r_{\rm{max}}=2$.
				(b) Fidelity (the blue vertical axis on the left side) of the target state versus $r_{\rm{max}}$. The red vertical axis on the right side denotes the mean photon number $\bar{n}_{\rm{d}}$ (red-dotted curve with marks ``+'') of the
				target state. The yellow-shaded area in (b) shows $\bar{n}_{\rm{d}}\geq 4$,
				indicating that the target state in this area can be called a large-size entangled cat state. To generate the large-size entangled cat state with $F \gtrsim 99\%$ and $\bar{n}_{\rm{d}}\simeq 4.3$, one can choose $r_{\rm{max}}= 2.3$ (the purple circle).
			}
		}
		\label{fig1}
	\end{figure}

	In the above numerical calculation of Fig.~\ref{fig0}(b), we have
	used the parameter
	${r}(t)={r}_{\rm{max}}/\{1+\exp{[{f}(t)]}\}$ with $f(t)=f_{0}\cos{(2\pi t/T)}$,
	where $f_{0}\gg 1$
	controls the initial and final values of the squeezing parameter $r(t)$.
	With these parameters,
	the pulses $\Omega_{\rm{r}}(t)$
	and $\Omega_{\rm{i}}(t)$ have finite durations, so that we can
	smoothly turn off the parametric drive [see Fig.~\ref{fig1}(a)].

	{In Fig.~\ref{fig1}(b), we show 
		the desired mean photon number $\bar{n}_{\rm{d}}=\langle G| a^{\dag}a|G\rangle$ versus the peak squeezing parameter $r_{\rm{max}}$ (red-dotted curve with ``+'').
		We find that, for a fixed evolution time $T$, $\bar{n}_{\rm{d}}$ increases sharply when $r_{\rm{max}}$ increases.
		Experimentally, a parametric gain of $\sim 20$~dB ($r_{\rm{max}}\sim 2.3$) has been achieved \cite{Sci3641163}, and $\sim 30$~dB has also been predicted under experimentally feasible conditions
		\cite{Nat541191,Nat49962,Prl121173601}.
		These realistic parameters allow for generating a high-fidelity ($F>90\%$) target state with $\bar{n}_{\rm{d}}=4\sim 10$ (large-amplitude nonclassical states),
		as shown by the blue, solid curve in Fig.~\ref{fig1}(b).
		Here, the fidelity of the state $|G\rangle$ is defined as $F=|\langle
		G|\rho(t_{f})|G\rangle|$.
		When $r_{\rm{max}}= 2.3$ and $\lambda=0.045\Delta$, we find that the
		target state $|G\rangle$ can be generated with $F\simeq 99\%$ and $\bar{n}_{\rm{d}}\simeq 4$
		[see purple circle in Fig.~\ref{fig1}(b)].
	}

\begin{figure}
		\centering
		\scalebox{0.3}{\includegraphics{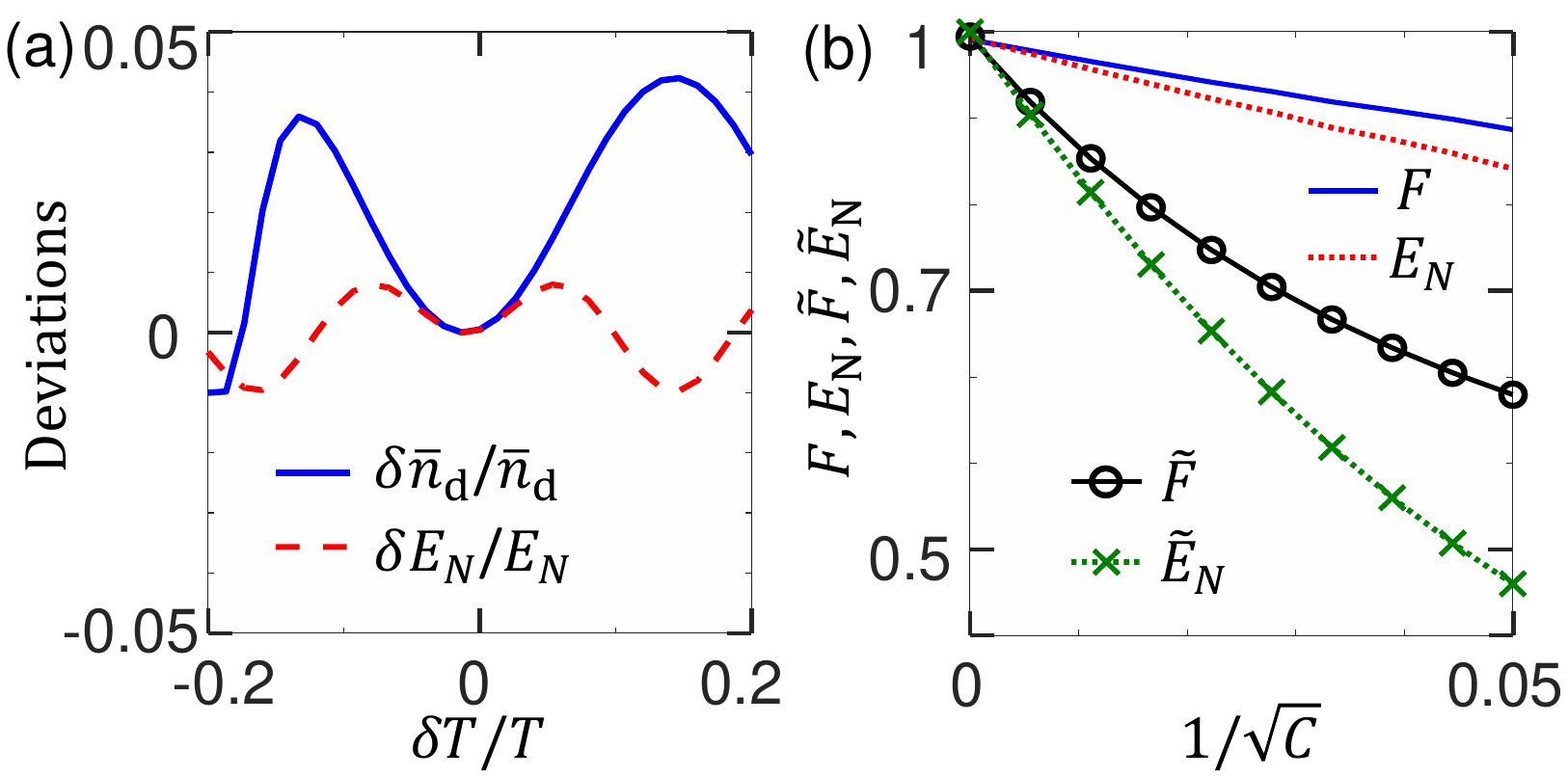}}
		\caption{(a) Deviations $\delta \bar{n}_{\rm{d}}$ and $\delta E_{N}$ versus $\delta T$.
			The STA protocol is robust against the imperfection of the total evolution time $T$.
			Parameters are the same as in Fig.~\ref{fig1}(a).
			(b) Fidelity $F$ ($\tilde{F}$) and logarithmic negativity $E_{N}$ ($\tilde{E}_{N}$) of the STA (adiabatic)
			protocol versus $1/\sqrt{C}$.
			Here, $C=\lambda^{2}/\kappa\gamma$ is
			the cooperativity, and we assume the dissipation rates $\gamma=\kappa$ for simplicity.
			The STA and the adiabatic protocols are initially equivalent by assuming $\lambda=\tilde{\lambda}=0.045\Delta$.
			The total evolution times for the entanglement generation are $T=20/\Delta$ and $\tilde{T}=250/\Delta$, respectively.
			The squeezing-induced noise is minimized by coupling the cavity to the squeezed-vacuum reservoir \cite{SM}.
			As a result, our STA protocol is much more robust against dissipation than the adiabatic protocol.
		}
		\label{fig3}
	\end{figure}

	\textit{Robustness of the STA approach}---In the following, we focus on discussing the robustness of the STA protocol when $r_{\max}=2.3$ and $\lambda=0.045\Delta$.
	We first assume
	the imperfection of a parameter $*$ as $\delta *=*'-*$, where $*'$ and $*$ denote the
	actual and ideal values, respectively. Because of large detuning $\lambda\ll\Delta$, when the parametric drive vanishes, the mean photon number and the entanglement of the system can remain unchanged for a long time
	in the absence of dissipation. Thus, our STA protocol is
	robust against the imperfect parameters of the total evolution time.
	{As shown in Fig.~\ref{fig3}(a), a $20\%$ imperfection of the total evolution time
		only causes $\lesssim1\%$ and $\lesssim 5\%$ changes of the logarithmic negativity $E_{N}$ and the mean photon number $\bar{n}_{\rm{d}}$, respectively.}
	
	Then, we compare the
	entanglement preparation via the STA and the adiabatic processes
	in the presence of cavity
	and qubit losses. {Because of the relatively strong squeezing,
		the difference of the frequencies of the photons ($\omega_{p}/2$) pumped by the two-photon drives and
		those of the squeezed-cavity mode ($\omega_{p}/2+\Delta\text{sech}[2(r(t))]$) is very small, such that
		the influence of the quantum fluctuation of the photons cannot be ignored.
		Thus, the two-photon drives may effectively excite the squeezed-cavity mode,
		so as to induce
		thermal noise and two-photon correlation noise in the squeezed-cavity mode \cite{Prl120093601,Prl120093602,Prl114093602,Pra100012339}.
		To minimize the influence of such noises, besides accelerating the dynamical evolution \cite{Q4271,Sci3641163},
		one can couple the cavity to a squeezed-vacuum reservoir \cite{Oe2113572,Oe2428383,Prl118103601} with $r_{e}=r_{\rm{max}}$
		and $\varphi_{e}=\pi$ during $T/4\lesssim t \lesssim 3T/4$ \cite{SM}.} Here $r_{e}$ and $\varphi_{e}$ are the squeezing parameter and the reference phase of the reservoir, respectively.
	In this case,
	the dynamics in the squeezed-light frame can be approximatively described by the standard Lindblad
	master equation
	\begin{align}\label{eq1-11}
	\dot{\rho}_{S}(t)\approx i[\rho_{S}(t),H_{S}(t)]+\gamma\mathcal{D}[\sigma]\rho_{S}(t)+\kappa\mathcal{D}[a]\rho_{S}(t),
	\end{align}
	where $\mathcal{D}[o]\rho_{S}(t)=o\rho_{S}(t) o^{\dag}-\left[o^{\dag}o \rho_{S}(t)+\rho_{S}(t)o^{\dag}o\right]/2$ is the standard Lindblad superoperator, $\rho_{S}(t)=S^{\dag}(t)\rho(t)S(t)$ is the density operator in the squeezed-light frame,
	$\gamma$
	is the spontaneous emission rate of the qubit, and $\kappa$ is the cavity decay rate.

	We define the
	cooperativity as $C=\lambda^2/\kappa\gamma$ and assume $\kappa\simeq\gamma$ for simplicity.
	By considering the same initial
	parameters $\lambda=\tilde{\lambda}=0.045\Delta$ and $r(0)=\tilde{r}(0)=0$,
	we compare the
	robustness of the STA and that of the adiabatic protocols [see Fig.~\ref{fig3}(b)] \cite{Qutip1}.
	The STA protocol is much more robust against
	dissipation than the adiabatic scheme, because: (i) the evolution time
	is significantly shortened in the STA protocol; (ii) the squeezing-induced noise can be well reduced by coupling the cavity to the squeezed-vacuum reservoir in the STA protocol.
	For experimentally realistic cavity QED parameters:
	$\Delta/2\pi=1$~GHz, $\lambda/2\pi=45$~MHz, and $\kappa/2\pi=\gamma/2\pi=2.25$~MHz,
	the STA protocol can achieve the target state with $F\sim 90\%$ and $E_{N}\sim 85\%$, while the adiabatic protocol fails ($\tilde{F}\sim 60\%$ and $\tilde{E}_{N}\sim 45\%$).
	Then, by measuring the qubit, we can achieve high-fidelity cat states in the lab frame.

	\textit{Conclusion.}---We have investigated how to simulate the STA dynamics of a cavity QED system in
	the strong coupling regime ($\lambda>\kappa,\gamma$)
	to prepare a maximally entangled cat state in the lab frame via parametric amplification.
	A significantly accelerated dynamics ($\sim10$ times faster than its adiabatic counterpart) makes
	the system much robust against dissipation.
	The target state is prepared in a large-detuned JC model, which is driven by finite-duration parametric pulses.
	Such a setup makes our STA protocol robust against the imperfection of the evolution time.
	Our proposal is feasible in circuit QED systems, where a transmission line resonator cavity interacts with a superconducting qubit in the JC model \cite{Np6772,Prl105237001,Pr7181,Rmp85623}. By attaching a superconducting quantum interference device (SQUID) to the end of the resonator \cite{Rmp841,Prb74224506,Prapplied12064037}, one can realize a two-photon drive (the Josephson parametric amplification process) by modulating in time the flux through the SQUID. \cite{Nc711338,Pra90053833,Prl103147003,Np4929,Apl91083509,Pra82052509,Eqt11,Nap94853}.
	{The squeezed vacuum (reservoir) is also produced by Josephson parametric amplifiers, but with
		a much larger linewidth than
		that of the cavity \cite{Prb83134501,Eqt11,Apl91083509,Apl107262601,Prapplied11034035,Nat49962,Prx6031004}}.
	This is possibly the first application of the STA protocols
	for the Rabi model and we hope that our protocol can find wide applications
	in studying light-matter interactions, specially, for the ultrastrong and deep-strong coupling regimes \cite{Nr119,Rmp91025005}.
	
	\begin{acknowledgements}
		We acknowledge helpful discussions with Yan-Xiong Du. Y.-H.C. is supported by the Japan Society for the Promotion of Science (JSPS) KAKENHI Grant No.~JP19F19028.
		X.W. is supported by the China
		Postdoctoral Science Foundation No.~2018M631136,
		and the Natural Science Foundation of China under
		Grant No.~11804270. A.M. is supported by the Polish National Science Centre (NCN)
		under the Maestro Grant No. DEC-2019/34/A/ST2/00081.
		F.N. is supported in part by: NTT Research, Army Research Office (ARO) (Grant No.~W911NF-18-1-0358), 
		Japan Science and Technology Agency (JST) (via the Q-LEAP program and the CREST Grant No.~JPMJCR1676), 
		Japan Society for the Promotion of Science (JSPS) (via the KAKENHI Grant No. JP20H00134 
		and the JSPS-RFBR Grant No. JPJSBP120194828), 
		the Asian Office of Aerospace Research and Development (AOARD), 
		and the Foundational Questions Institute Fund (FQXi) via Grant No. FQXi-IAF19-06.
	\end{acknowledgements}

	\bibliography{references}

\newpage
\clearpage
\appendix

\setcounter{equation}{0} \setcounter{figure}{0}
\setcounter{table}{0} \setcounter{page}{1} \makeatletter
\renewcommand{\theequation}{S\arabic{equation}}
\renewcommand{\thefigure}{S\arabic{figure}}
\renewcommand{\bibnumfmt}[1]{[S#1]}
\renewcommand\thesection{S\arabic{section}}
\renewcommand{\thetable}{S\arabic{table}}
\begin{widetext}
\section{Supplemental Material}
{In this Supplemental Material, we first
	discuss the influence of the nonadiabatic transition caused by mapping the system dynamics into the time-dependent squeezed-light frame. Using the transitionless algorithm, we show how
	to counteract such a nonadiabatic transition with an additional drive, so as to design
	a shortcuts-to-adiabatic passage to rapidly generate giant entangled cat states.
	{Then, we show how to minimize the influence of the squeezing-induced fluctuation noise by coupling the cavity to a squeezed-vacuum reservoir.}
	Thirdly, we present a possible
	problem in turning off the parametric drive when the target
	state is generated in the squeezed-light frame via the adiabatic process.}

\section{S1.~Effective Hamiltonian and Dissipation dynamics of the system}
\subsection{A.~Counteracting the nonadiabatic transition caused by the time-dependent unitary transformation}

We begin with a largely detuned Jaynes-Cummings (JC) Hamiltonian driven by a time-dependent parametric (two-photon) drive $\Omega_{r}(t)$,
\begin{align}\label{S1}
{H}_{\rm{1}}(t)=\Delta{a}^{\dag}{a}
-\left[\frac{\Omega_{r}(t)}{2}{a}^{2}-\lambda a^{\dag}\sigma+\text{h.c.}\right].
\end{align}
In the time-dependent squeezed-light frame determined by the squeezing operator $S(t)=\exp{[r(t)({a}^{\dag 2}-{a}^{2})/2]}$,
with a real squeezing parameter $r(t)$ satisfying $\tanh\left[2r(t)\right]=\Omega_{r}(t)/\Delta$,
the Hamiltonian of the system is composed of the following terms:
\begin{align}\label{S2}
H_{\rm{S1}}(t)&=S^{\dag}(t)H_{1}(t)S(t)-iS^{\dag}(t)\dot{S}(t) \cr
&={H}_{\rm{S-Rabi}}(t)+H_{\rm{err}}(t)+H_{\rm{NA}}(t), \cr
H_{\rm{S-Rabi}}(t)&=\Delta\text{sech}{[2r(t)]}a^{\dag}a+\lambda{\exp{[r(t)]}}{\sigma_{x}}( a^{\dag}+a)/2, \cr
H_{\rm{err}}(t)&=-i\lambda{\exp{[-r(t)]}}\sigma_{y}(a^{\dag}-a)/2, \cr
H_{\rm{NA}}(t)&=-i\dot{r}(t)(a^{\dag 2}-a^{2})/2.
\end{align}
The Hamiltonian $H_{\rm {S-Rabi}}$ describes
the $\sigma_x X$ Rabi interaction in the squeezed-light frame,
where $X=(a+a^{\dag})/2$ is the canonical position operator.
The Hamiltonian $H_{\rm{err}}(t)$ describes the $\sigma_{y} Y$ interaction,
where $Y=i(a^{\dag}-a)/2$ is the canonical momentum operator, and can
be considered an error term, which
can be neglected when $\lambda\ll \Delta$ and $\lambda/\Delta\ll r(t)$.
When $r(t)\sim \lambda/\Delta$, the error term $H_{\rm{err}}(t)$ can be neglected by applying a strong drive $\Omega\sigma_{x}$ ($\Omega\gtrsim\Delta$),
which induces the coupling of $H_{\rm{err}}(t)$ with a large detuning in the $\sigma_{y}$-direction.

The last term in $H_{\rm{S1}}(t)$, i.e., $H_{\rm{NA}}(t)=-iS^{\dag}(t)\dot{S}(t)$, describes a nonadiabatic transition induced by mapping the system dynamics into the time-dependent squeezed-light frame. It describes the population transfer
between different basis in the squeezed-light frame.
According to Berry's transitionless algorithm, we can add a
term
\begin{align}\label{S3}
H_{\rm{SA}}(t)=iS^{\dag}(t)\dot{S}(t)=i\dot{r}(t)(a^{\dag 2}-a^{2})/2,
\end{align}
into the Hamiltonian $H_{\rm{S1}}(t)$ to counteract the nonadiabatic transition.
Then, in the laboratory frame, the additional Hamiltonian $H_{\rm{SA}}$ reads
\begin{align}\label{S4}
H_{\rm{add}}(t)=S(t)H_{\rm{SA}}(t)S^{\dag}=i\dot{r}(t)(a^{\dag 2}-a^{2})/2.
\end{align}
This implies that the cavity mode is subject to another two-photon drive, which has an amplitude $\Omega_{i}(t)=\dot{r}(t)$,
a frequency $\omega_{{p}}$, and is $\pi/2$-dephased from $\Omega_{r}(t)$.
By adding this additional Hamiltonian $H_{\rm{add}}(t)$ into the Hamiltonian $H_{1}(t)$, we obtain the
Hamiltonian $H_{0}(t)$ required for the STA protocol, i.e., the Hamiltonian of Eq.~(6) of the main text:
\begin{align}\label{S5}
{H}_{0}(t)=\Delta{a}^{\dag}{a}+\Omega\sigma_{x}
-\left[\frac{\Omega_{r}(t)+i\Omega_{i}(t)}{2}{a}^{2}-\lambda{a^{\dag}}\sigma+\text{h.c.}\right].
\end{align}
Then, we are allowed to rapidly change the squeezing parameter $r(t)$, such
that we can quickly adjust the effective qubit-cavity coupling $\lambda \exp{[r(t)]}/2$ in the squeezed-light frame.

This is very important, because applying the STA protocol requires to rapidly change the control parameter, i.e., the normalized coupling strength.

{\subsection{B.~STA process with parametric drivings}
	To construct the STA passage, we divide the Hamiltonian $H_{S}(t)$ into two parts:
	\begin{align}\label{R6}
	H_{S}(t)=H_{\rm{ref}}(t)+H_{\rm{aux}}(t).
	\end{align}
	Here, the Hamiltonian
	\begin{align}\label{R7}
	H_{\rm{ref}}(t)=\Delta\text{sech}{[2r(t)]}a^{\dag}a+\sigma_{x}[\chi(t) a^{\dag}+\chi^{*}(t)a],
	\end{align}
	is considered as the reference Hamiltonian [with an undetermined parameter $\chi(t)$] for constructing shortcuts, 
	\begin{align}\label{R8}
	H_{\rm{aux}}(t)=\frac{\lambda{e^{r(t)}}}{2}{\sigma_{x}}( a^{\dag}+a)-\sigma_{x}[\chi(t) a^{\dag}+\chi^{*}(t)a],
	\end{align}
	is an auxiliary Hamiltonian. The reference Hamiltonian $H_{\rm{ref}}(t)$ takes the same form 
	as the Rabi Hamiltonian $H_{R}(t)$ [Eq.~(1) of the main text], i.e., $H_{\rm{ref}}(t):\Rightarrow H_{R}(t)$, by setting:
	\begin{align}
	\omega_{{q}}\ll \omega_{{c}},\ \ \ \omega_{{c}}:\Rightarrow\Delta\text{sech}{[2r(t)]},\ \ \  g:\Rightarrow\chi(t).
	\end{align}
	Then, when we choose the parameters to satisfy 
	\begin{align}\label{R9}
	\eta(t)=\frac{g}{\omega_{c}}=\frac{\chi(t)}{\Delta\text{sech}[2r(t)]},\ \ \ \ \dot{\eta}(t)=\frac{i}{2}\left[{\lambda{e^{r(t)}}}-2\chi(t)\right],
	\end{align}
	$H_{\rm{aux}}(t)$ is exactly the CD driving Hamiltonian for the reference Hamiltonian $H_{\rm{ref}}(t)$, i.e., $H_{\rm{aux}}(t):\Rightarrow H_{\rm{CD}}(t)$.
	Hence, according 
	to the transitionless algorithm, the CD driving Hamiltonian $H_{\rm{aux}}(t)$ can actually drive the system to evolve along an eigenstate of $H_{\rm{ref}}(t)$. The evolution path for our STA protocol is then given as (in the squeezed-light frame)
	\begin{align}\label{R10}
	|E_{0}(t)\rangle_{S}=\frac{1}{\sqrt{2}}\left[|+_{x}\rangle|-{\eta}(t)\rangle+|-_{x}\rangle|{\eta}(t)\rangle\right],
	\end{align}
	where $|\pm_{x}\rangle$ are the eigenstates of the Pauli matrix $\sigma_{x}$.
	In the lab frame, the STA evolution path is $S[r(t)]|E_{0}(t)\rangle_{S}$.
	After some algebra, we can counteract the undetermined parameter $\chi(t)$ and obtain the equations of motion for the coherent state amplitude $\eta(t)$:
	\begin{align}\label{R11}
	\text{Re}[\dot{\eta}(t)]=&\Delta\text{Im}[\eta(t)]\text{sech}2r(t),\cr
	\text{Im}[\dot{\eta}(t)]=&\frac{\lambda}{2} \exp[{r(t)}]-\Delta\text{Re}[\eta(t)]\text{sech}2r(t).
	\end{align}
	Thus, Eq.~(9) of the main text is obtained.
	The final state in the laboratory frame is 
	\begin{align}\label{S10}
	S(t_{f})|E_{0}(t_{f})\rangle_{S}=\frac{1}{\sqrt{2}}\left(|+_{x}\rangle|-{\eta}(t_{f})\rangle+|-_{x}\rangle|{\eta}(t_{f})\rangle\right),
	\end{align}
	which is an entangled cat state. Here, $S{(t_{f})}=1$ is given according to $r(t_{f})=0$.

}

\subsection{C.~Minimizing the influence of the squeezing-induced fluctuation noise}
The Markovian master equation, for a cavity interacting
with a broadband squeezed-vacuum reservoir (at zero temperature with squeezing parameter $r_{e}$ and reference phase $\varphi_{e}$),
has been well studied  (see, e.g., Ref.~\cite{Scully1997}). For our STA protocol, when the cavity couples to
the squeezed-vacuum reservoir, the master equation in the laboratory frame is
\begin{align}\label{S3-1}
\dot{\rho}(t)=&i[\rho(t),H_{0}(t)]+\frac{1}{2}\left[2L_{\gamma}\rho(t)L^{\dag}_{\gamma}
-L_{\gamma}^{\dag}L_{\gamma}\rho(t)-\rho(t)L_{\gamma}^{\dag}L_{\gamma}\right] \cr
&+\frac{1}{2}(N+1)\left[2L_{\kappa}\rho(t)L^{\dag}_{\kappa}
-L_{\kappa}^{\dag}L_{\kappa}\rho(t)-\rho(t)L_{\kappa}^{\dag}L_{\kappa}\right] \cr
&+ \frac{1}{2}N\left[2L_{\kappa}^{\dag}\rho(t)L_{\kappa}
-L_{\kappa}L_{\kappa}^{\dag}\rho(t)-\rho(t)L_{\kappa}L_{\kappa}^{\dag}\right] \cr
&-\frac{1}{2}M\left[2L_{\kappa}^{\dag}\rho(t)L_{\kappa}^{\dag}
-L_{\kappa}^{\dag}L_{\kappa}^{\dag}\rho(t)-\rho(t)L_{\kappa}^{\dag}L_{\kappa}^{\dag}\right] \cr
&-\frac{1}{2}M^{*}\left[2L_{\kappa}\rho(t)L_{\kappa}
-L_{\kappa}L_{\kappa}\rho(t)-\rho(t)L_{\kappa}L_{\kappa}\right].
\end{align}
Here, $L_{\gamma}=\sqrt{\gamma}\sigma$ and $L_{\kappa}=\sqrt{\kappa}a$ describe the qubit and cavity decays, with decay rates $\gamma$ and $\kappa$, respectively. The parameters
\begin{align}\label{S3-2}
N=\sinh^{2}(r_{e}), \ \ \ \ \text{and} \ \ \ \ M=\cosh{(r_{e})}\sinh{(r_{e})}\exp{(-i\varphi_{e})},
\end{align}
describe thermal noise and two-photon correlation noise caused by the squeezed-vacuum reservoir, respectively.

\begin{figure}
	\centering
	\scalebox{0.65}{\includegraphics{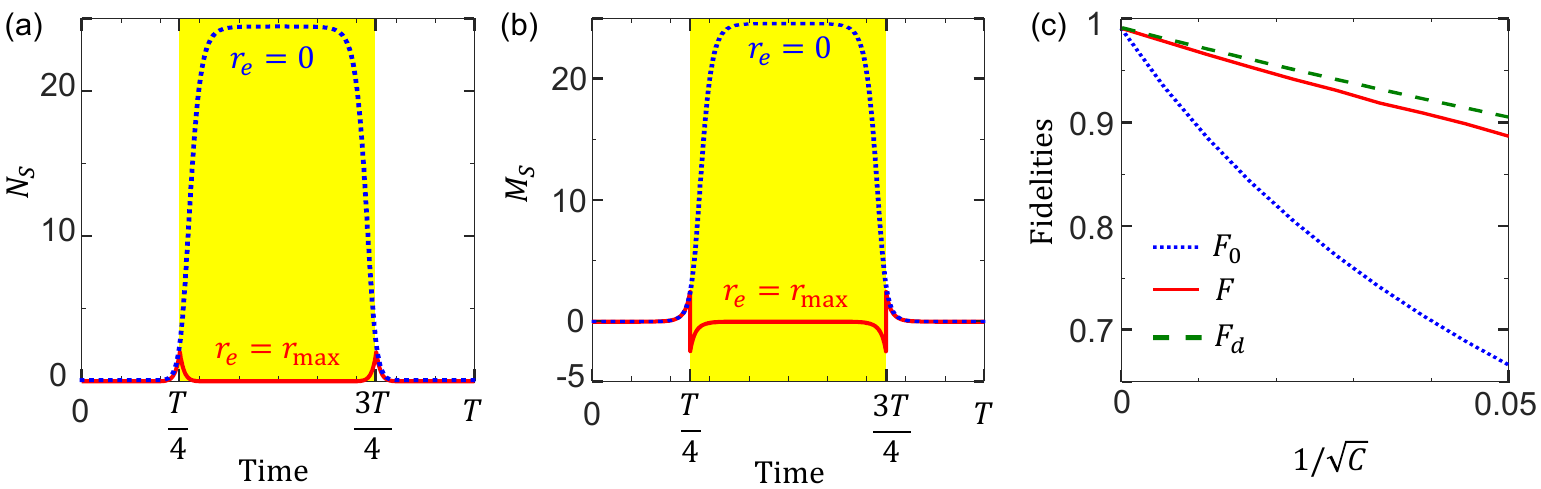}}
	\caption{Parameters (a) $N_{S}$ and (b) $M_{S}$ characterizing the squeezing-induced noise for $r_{\rm{max}}=2.3$ and $\Omega=0$. Blue-dotted curve in (a) [(b)]: Parameter $N_{S}$ ($M_{S}$) without coupling the cavity to the squeezed-vacuum reservoir (i.e., $r_{e}=0$); Red-solid curve in (a) [(b)]: Parameter $N_{S}$ ($M_{S}$) when the system is coupled to the squeezed-vacuum reservoir during $T/4\lesssim t \lesssim 3T/4$ [i.e., $r_{e}$ is given according to Eq.~(\ref{S3-7})]. The yellow-shaded area in (a) or (b) denotes when the cavity is coupled to the squeezed-vacuum reservoir.
		(c) Fidelities of the ground state $|G\rangle$ versus $1/\sqrt{C}$ calculated by: (blue-dotted curve representing ${F}_{\rm{0}}$)
		the noise-included master equation in Eq.~(\ref{S3-3}) when $r_{e}=0$;
		(red-solid curve representing ${F}$) the noise-included master equation when coupling the cavity to the squeezed-vacuum;  (green-dashed curve representing ${F}_{d}$) the effective master equation in Eq.~(\ref{S3-7c}).	
		The parameter $C=\lambda^{2}/\kappa\gamma$ is
		the cooperativity, and we assume the dissipation rates $\gamma=\kappa$ for simplicity.
	}
	\label{figS2}
\end{figure}

By mapping the system dynamics into the time-dependent squeezed-light frame with $S(t)$, the master equation becomes
\begin{align}\label{S3-3}
\dot{\rho_{S}}(t)=&i[\rho_{S}(t),H_{\rm{S-Rabi}}(t)+H_{\rm{err}}(t)]+\frac{1}{2}\left[2L_{\gamma}\rho_{S}(t)L^{\dag}_{\gamma}
-L_{\gamma}^{\dag}L_{\gamma}\rho_{S}(t)-\rho_{S}(t)L_{\gamma}^{\dag}L_{\gamma}\right] \cr
&+\frac{1}{2}(N_{S}+1)\left[2L_{\kappa}\rho_{S}(t)L^{\dag}_{\kappa}
-L_{\kappa}^{\dag}L_{\kappa}\rho_{S}(t)-\rho_{S}(t)L_{\kappa}^{\dag}L_{\kappa}\right] \cr
&+ \frac{1}{2}N_{S}\left[2L_{\kappa}^{\dag}\rho_{S}(t)L_{\kappa}
-L_{\kappa}L_{\kappa}^{\dag}\rho_{S}(t)-\rho_{S}(t)L_{\kappa}L_{\kappa}^{\dag}\right] \cr
&-\frac{1}{2}M_{S}\left[2L_{\kappa}^{\dag}\rho_{S}(t)L_{\kappa}^{\dag}
-L_{\kappa}^{\dag}L_{\kappa}^{\dag}\rho_{S}(t)-\rho_{S}(t)L_{\kappa}^{\dag}L_{\kappa}^{\dag}\right] \cr
&-\frac{1}{2}M_{S}^{*}\left[2L_{\kappa}\rho_{S}(t)L_{\kappa}
-L_{\kappa}L_{\kappa}\rho_{S}(t)-\rho_{S}(t)L_{\kappa}L_{\kappa}\right],
\end{align}
where $\rho_{S}(t)=S^{\dag}(t)\rho(t)S(t)$ is the density operator of the system in the squeezed-light frame, and
\begin{align}\label{S3-4}
N_{S}=&\cosh^{2}[r(t)]\sinh^{2}(r_{e})+\sinh^{2}[r(t)]\cosh^{2}(r_{e})+\frac{1}{2}\sinh[2r(t)]\sinh(2 r_{e})\cos(\varphi_{e}),  \cr
M_{S}=&\left\{\sinh[r(t)]\cosh(r_{e})+\exp{(-i\varphi_{e})}\cosh[r(t)]\sinh(r_{e})\right\}\cr &\times\left\{\cosh[r(t)]\cosh(r_{e})+\exp(i\varphi_{e})\sinh[r(t)]\sinh(r_{e})\right\},
\end{align}
characterize additional noises of the system in the squeezed-light frame.
When $r_{e}=0$, $N_{S}$ and $M_{S}$ characterize the squeezing-induced noise. For simplicity, we can assume $\varphi_{e}=\pi$,
and obtain
\begin{align}\label{S3-5a}
N_{S}=\sinh^{2}\left[r_{S}(t)\right], \ \ \ \text{and} \ \ \ M_{S}=\cosh\left[r_{S}(t)\right]\sinh\left[r_{S}(t)\right],
\end{align}
where $r_{S}(t)=r(t)-r_{e}$. Then, to minimize the parameters $|N_{S}|$ and $|M_{S}|$, we need to minimize the parameter $|r_{S}(t)|$.

\begin{figure}
	\centering
	\scalebox{0.6}{\includegraphics{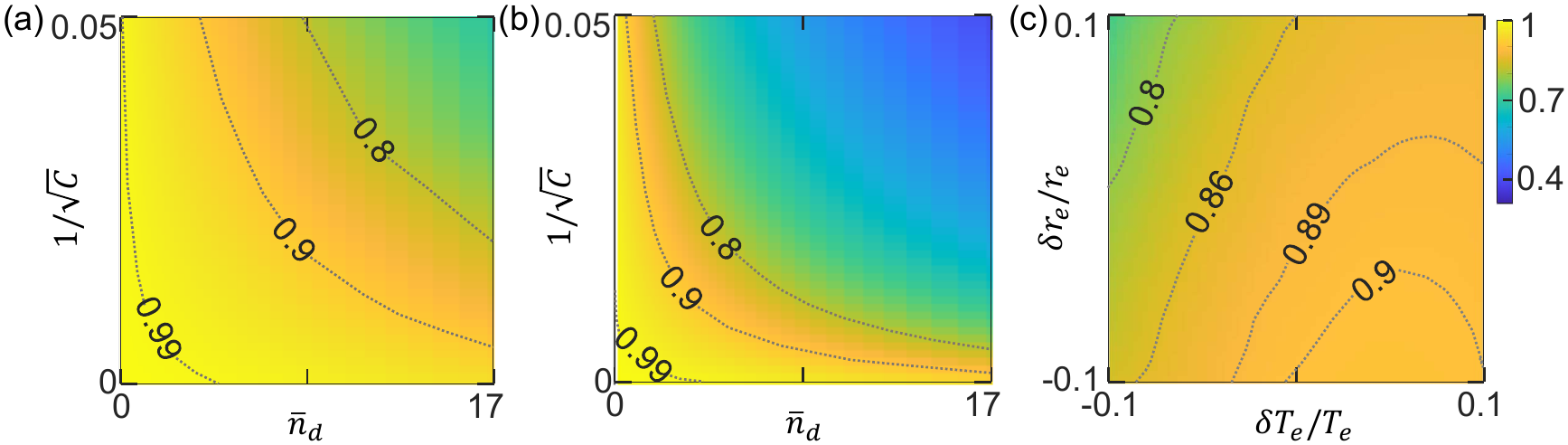}}
	\caption{{{blue} Fidelities $F$ of the entangled cat state for the STA protocol with
			$T=20/\Delta$ and $\lambda=0.045\Delta$: (a) [(b)] Fidelity $F$ versus
			the cooperativity $C$ and the predicted mean photon number $\bar{n}_{d}=|\eta|^{2}$, when the cavity is coupled (decoupled) to
			the squeezed-vacuum reservoir.
			For the panels (a) and  (b), we choose the squeezing parameter $r_{\rm{max}}\in[0,3]$ 
			corresponding to the $\bar{n}_{d}=|\eta|^{2}\in[0,17]$. (c) Fidelity $F$ versus 
			imperfections of the parameters $r_{e}$ and $T_{e}$. For the panel (c),
			the squeezing parameter is $r_{\rm{max}}=2.3$.}
	}
	\label{RF2}
\end{figure}

The waveform of $r\left(t\right)$ of the STA protocol is approximately a square wave when
\begin{align}\label{S3-5}
{r}(t)=\frac{{r}_{\rm{max}}}{1+\exp{[f_{0}\cos{(2\pi t/T)}]}},
\end{align}
where $f_{0}=10$ controls the initial and final values of the squeezing parameter $r(t)$. Substituting Eq.~(\ref{S3-5}) into Eq.~(\ref{S3-5a}) and assuming $r_{e}=0$, in Figs.~\ref{figS2}(a) and \ref{figS2}(b), we show the parameters $N_{S}$ and $M_{S}$ describing the squeezing-induced noise (see the blue-dotted curves). As shown, the squeezing-induced noise affects the system dynamics especially when $r(t)$ reaches its maximum value $r_{\rm{max}}$, i.e., $r(t)\approx r_{\rm{max}}$. We accordingly calculate the fidelity $F_{0}=|\langle G|\rho_{S}(t_{f})|G\rangle|$ to show the
influence of the squeezing-induced noise [see the blue-dotted curve in
Fig.~\ref{figS2}(c)]. Here, $|G\rangle$ is the ground state of the Rabi model in the DSC regime
[see Eq.~(2) of the main text]. The fidelity $F_{0}$ decreases very fast when the
dissipation increases.

To minimize the parameter $|r_{S}(t)|$, according to the
properties of $\cos(2\pi t/T)$, we can choose
\begin{align}\label{S3-6}
{r}_{e}=\left\{
\begin{array}{rcl}
0,& \ \ \ & (0\lesssim t \lesssim T/4) \\ \\
r_{\rm{max}},&  & (T/4\lesssim t \lesssim 3T/4) \\ \\
0,& &  (3T/4\lesssim t \lesssim T)
\end{array}
\right.
\end{align}
i.e., the total interaction time between the cavity and the squeezed-vacuum reservoir is $T_{e}=T/2$,
resulting in
\begin{align}\label{S3-7}
{r}_{S}(t)=\left\{
\begin{array}{rcl}
\displaystyle\frac{{r}_{\rm{max}}}{1+\exp{[f_{0}\cos{(2\pi t/T)}]}}, & \ \ \ & (0\lesssim t \lesssim T/4) \\ \\
\displaystyle\frac{-{r}_{\rm{max}}}{1+\exp{[-f_{0}\cos{(2\pi t/T)}]}},&  & (T/4\lesssim t \lesssim 3T/4) \\ \\
\displaystyle\frac{{r}_{\rm{max}}}{1+\exp{[f_{0}\cos{(2\pi t/T)}]}}.& &  (3T/4\lesssim t \lesssim T)
\end{array}
\right.
\end{align}
Then, substituting Eq.~(\ref{S3-7}) into Eq.~(\ref{S3-5a}),
we plot the parameters $N_{S}$ and $M_{S}$ [see
the red-solid curves in Fig.~\ref{figS2}(a) and \ref{figS2}(b)].
We can accordingly calculate the average values
\begin{align}\label{S3-7b}
A_{N_{S}}=\frac{1}{T}\int_{0}^{t_{f}}|N_{S}|dt\approx 0.08,\ \ \ \text{and} \ \ \
A_{M_{S}}=\frac{1}{T}\int_{0}^{t_{f}}|M_{S}|dt\approx 0.14 ,
\end{align}
which means that the additional noises in Eq.~(\ref{S3-3}) weakly affect the system dynamics. Thus, the fidelity of the target state $|G\rangle$ is  significantly improved [see the red-solid curve in Fig.~\ref{figS2}(c)], e.g., from $\sim 65\%$ to $ \sim 89\%$ when $1/\sqrt{C}=0.05$.
{ When the desired mean photon number $\bar{n}_{d}$ of the target state increases,
	the influence of the cavity loss increases [see Figs.~\ref{figS3}(a) and \ref{figS3}(b)].
	These figures show the fidelities of the
	target state when the cavity is coupled and decoupled 
	to the squeezed-vacuum reservoir, respectively. According to the comparison between
	Figs.~\ref{figS2}(a) and \ref{figS2}(b), coupling the cavity to the squeezed-vacuum reservoir can effectively suppress the influence of the
	cavity loss. Thus, the giant ($\bar{n}_{d}\gtrsim 10$) entangled cat states can be generated with a high fidelity. By defining the imperfection of a parameter $*$ as $\delta *=*'-*$,
	the influence of the imperfections of the parameters $T_{e}$ and $r_{e}$ is shown in
	Fig.~\ref{figS2}(c). This figure shows that,
	slightly decreasing the squeezing parameter $r_{e}$ or increasing the interaction time $T_{e}$ can improve the fidelity $F$. Note that a $10\%$ imperfection of the
	parameter $r_{e}$ only causes a $3\%$ change in the fidelity, thus the STA protocol
	is mostly insensitive to the imperfections of the
	parameter $r_{e}$.  When the interaction time $T_{e}$ between
	the cavity and the squeezed-vacuum reservoir is long enough, our STA protocol
	is mostly insensitive to the imperfections of the
	parameter $T_{e}$.
}

When coupling the cavity to the squeezed-vacuum reservoir during $T/4\lesssim t \lesssim 3T/4$, the evolution can be approximately described by the standard Lindblad master equation
\begin{align}\label{S3-7c}
\dot{\rho_{S}}(t)\approx&i[\rho_{S}(t),H_{\rm{S-Rabi}}(t)]+\frac{1}{2}\sum_{m=\kappa,\gamma}\left[2L_{m}\rho_{S}(t)L^{\dag}_{m}
-L_{m}^{\dag}L_{m}\rho_{S}(t)-\rho_{S}(t)L_{m}^{\dag}L_{m}\right],
\end{align}
which is Eq.~(10) of the main text. As shown in Fig.~\ref{figS2}(c),
the Lindblad master equation can well describe the dynamics when the cavity
is coupled to the squeezed-vacuum reservoir.

\begin{figure}
	\centering
	\scalebox{0.65}{\includegraphics{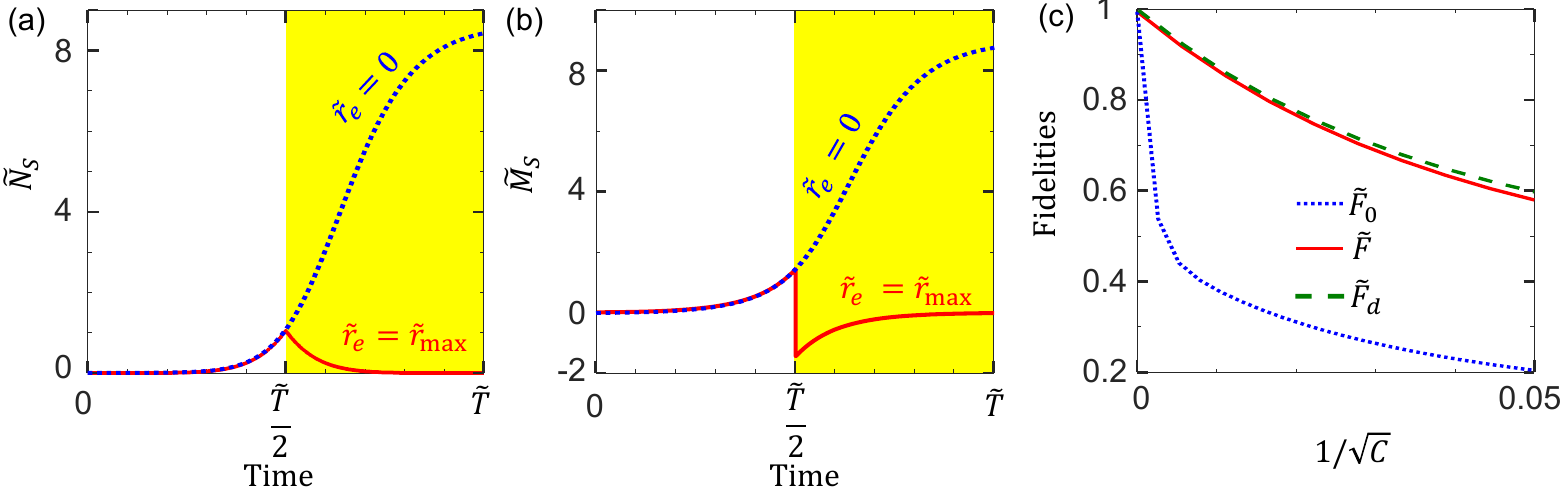}}
	\caption{Parameters (a) $\tilde{N}_{S}$ and (b) $\tilde{M}_{S}$  characterizing the squeezing-induced noise for $\tilde{r}_{\rm{max}}=1.8$. Blue-dotted curve in (a) [(b)]: parameter $\tilde{N}_{S}$ ($\tilde{M}_{S}$) without coupling the cavity to the squeezed-vacuum reservoir (i.e., $\tilde{r}_{e}=0$); Red-solid curve in (a) [(b)]: parameter $\tilde{N}_{S}$ ($\tilde{M}_{S}$) when the system is coupled to the squeezed-vacuum reservoir during $\tilde{T}/2\lesssim t \lesssim \tilde{T}$ [i.e., $\tilde{r}_{e}$ is given according to Eq.~(\ref{S3-9})]. The yellow-shaded area in (a) or (b) denotes that the cavity is coupled to the squeezed-vacuum reservoir. (c) Fidelities of the squeezed ground state $|SG\rangle=S(t_{f})|G\rangle$ versus $1/\sqrt{C}$ calculated by: (blue-dotted curve representing $\tilde{F}_{\rm{0}}$)
		the noise-included master equation in Eq.~(\ref{S3-3}) when $r_{e}=0$; (red-solid curve representing $\tilde{F}$) the noise-included master equation when coupling the cavity to the squeezed-vacuum reservoir; (green-dashed curve representing $\tilde{F}_{d}$) the effective master equation in Eq.~(\ref{S3-7c}).
	}
	\label{figS3}
\end{figure}

This strategy is also applicable in the adiabatic protocol to minimize the influence of the squeezing-induced noise.
For the adiabatic protocol, the squeezing parameter $\tilde{r}(t)$ is
\begin{align}\label{S3-8}
\tilde{r}(t)=\frac{\tilde{r}_{\rm{max}}}{1+\exp{[\tilde{f}_{0}(1/2-t/\tilde{T})]}},
\end{align}
where $\tilde{f}_{0}=10$ controls the initial and final values of $\tilde{r}(t)$.
Substituting Eq.~(\ref{S3-8}) into Eq.~(\ref{S3-4}) and assuming $\tilde{r}_{e}=0$, we plot the parameters $\tilde{N}_{S}$ and $\tilde{M}_{S}$ in Figs.~\ref{figS3}(a) and \ref{figS3}(b). We denote $\tilde{*}$ ($*=r,\ T,\ \ldots$) to represent the parameters used in the adiabatic protocol. The parameter $\tilde{*}$ has the same physical meaning as $*$. Due to the squeezing-induced noise, the adiabatic protocol
becomes unreliable for the finite cooperativity $C$ [see the blue-dotted curve in Fig.~\ref{figS3}(c)].

To minimize the parameters $|\tilde{N}_{S}|$ and $|\tilde{M}_{S}|$, we can assume
\begin{align}\label{S3-9}
\tilde{r}_{e}=\left\{
\begin{array}{rcl}
0,& \ \ \ & (0\lesssim t \lesssim \tilde{T}/2) \\ \\
\tilde{r}_{\rm{max}},&  & (\tilde{T}/2\lesssim t \lesssim \tilde{T})
\end{array}
\right.
\end{align}	
resulting in
\begin{align}\label{S3-10a}
\tilde{r}_{S}(t)=\left\{
\begin{array}{rcl}
\displaystyle\frac{\tilde{r}_{\rm{max}}}{1+\exp{[\tilde{f}_{0}(1/2-t/\tilde{T})]}},& \ \ \ & (0\lesssim t \lesssim \tilde{T}/2) \\ \\
\displaystyle\frac{-\tilde{r}_{\rm{max}}}{1+\exp{[-\tilde{f}_{0}(1/2-t/\tilde{T})]}}.&  & (\tilde{T}/2\lesssim t \lesssim \tilde{T})
\end{array}
\right.
\end{align}		
Accordingly, the average values of $|\tilde{N}_{S}|$ and $|\tilde{M}_{S}|$ are
\begin{align}\label{S3-10}
\tilde{A}_{N_{S}}=\frac{1}{\tilde{T}}\int_{0}^{t_{f}}|\tilde{N}_{S}|dt\approx 0.14,\ \ \ \text{and} \ \ \
\tilde{A}_{M_{S}}=\frac{1}{\tilde{T}}\int_{0}^{t_{f}}|\tilde{M}_{S}|dt\approx 0.3,
\end{align}
respectively.
Thus, the additional noises characterized by $\tilde{N}_{S}$
and $\tilde{M}_{S}$ can be suppressed as shown in Fig.~\ref{figS3}(a) and \ref{figS3}(b).
The fidelity of the squeezed ground state
$|SG\rangle=S(t_{f})|G\rangle$ is improved [see the red-solid curve in Fig.~\ref{figS3}(c)].
However, due to
\begin{align}\label{S3-11}
\tilde{A}_{N_{S}}> {A}_{N_{S}},\ \ \ \tilde{A}_{M_{S}}> {A}_{M_{S}},\ \ \ \text{and}\ \ \ \tilde{T}\gg T,
\end{align}
the squeezing-induced noise still affects the adiabatic protocol more seriously than the STA protocol. Thus, the fidelity of the adiabatic protocol is much lower than
the STA method, according to the comparison between Figs.~\ref{figS2}(c) and \ref{figS3}(c).

\begin{figure}
	\centering
	\scalebox{0.45}{\includegraphics{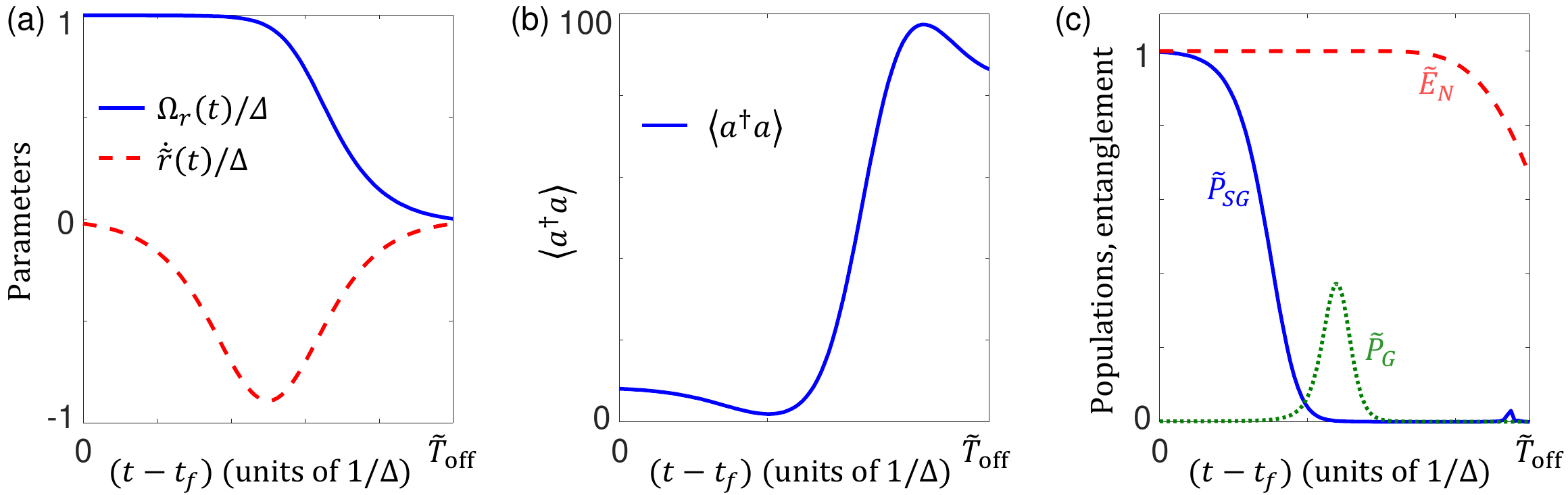}}
	\caption{System evolution during turning off the parametric drive in the adiabatic protocol for $t>t_{f}$.
		(a) Amplitude of the parametric drive $\Omega_{r}(t)$
		and the changing rate $\dot{\tilde{r}}(t)$ of the squeezing parameter $r(t)$.
		(b) Mean photon number $\langle a^{\dag} a\rangle$ of the system.
		(c) Blue-solid curve: the population $\tilde{P}_{{SG}}$ of the squeezed state $|SG\rangle$; Green-dotted curve: the population $\tilde{P}_{{G}}$ of the ground state $|G\rangle$; Red-dashed curve: the entanglement cost (characterized by the logarithmic negativity $\tilde{E}_{{N}}$).
		The time required to turn off the parametric drive is assumed to be $\tilde{T}_{\rm{off}}=5/\Delta$.
	}
	\label{figS1}
\end{figure}

\section{S2. A Possible Problem caused by turning off the parametric drive in the adiabatic protocol}
The nonadiabatic transition $H_{\rm{NA}}(t)$ also causes the main problem of how to turn off the parametric drive.
In the adiabatic protocol discussed in the main text, the amplitude of the parametric drive $\Omega_{r}(t)$
reaches the peak value at the time $t_{f}$, i.e., $\Omega_{r}(t_{f})=\Omega_{\rm{max}}$.
Meanwhile, the maximally entangled cat state is prepared in the squeezed
frame. In the laboratory frame, the final state corresponds to the qubit
being entangled with the squeezed and displaced cavity pointer states, i.e., $|SG\rangle$.
To smoothly and rapidly turn off
the parametric drive, we can assume
\begin{align}\label{S6}
\tilde{r}(t)=\frac{1}{2}\frac{\text{arctanh}(\Omega_{\rm{max}}/\Delta)}{{1+\exp{\{10[-(t-t_{f})/\tilde{T}_{\rm{off}}+1/3]\}}}},\ \ \ \ \ (t\geq t_{f})
\end{align}
corresponding to
\begin{align}\label{S7}
\tilde{r}(t_{f})=\frac{1}{2}\text{arctanh}(\Omega_{\rm{max}}/\Delta),\ \ \
\tilde{r}(t_{f}+\tilde{T}_{\rm{off}})\simeq 0,  \ \ \
\dot{\tilde{r}}(t_{f})\simeq 0, \ \ \
\dot{\tilde{r}}(t_{f}+\tilde{T}_{\rm{off}})\simeq 0.
\end{align}
Here, $\tilde{T}_{\rm{off}}$ is the operation time required to turn off the parametric drive.

Assuming $\tilde{T}_{\rm{off}}=5/\Delta$ as an example,
we show $\Omega_{r}(t)$ and $\dot{\tilde{r}}(t)$ versus time
in Fig.~\ref{figS1}(a).
Due to $\dot{\tilde{r}}(t)\neq 0$,
the nonadiabatic transition $H_{\rm{NA}}(t)$ can pump many photons
into the cavity.
By substituting Eq.~(\ref{S6}) into Eq.~(\ref{S2}), and assuming the
system is in the squeezed ground state $|SG\rangle$ at the time
$t_{f}$, we
show the instantaneous mean photon number $\langle a^{\dag}a\rangle$ when $t>t_{f}$ in
Fig.~\ref{figS1}(b). {We find that $\langle a^{\dag}a\rangle$ increases sharply when $\Omega_{r}(t)$ decreases. When the parametric drive is turned off,
	i.e., $\Omega_{r}(t)=0$, the desired entangled state does not exist any longer [see in Fig.~\ref{figS1}(c)].
	Both populations of the squeezed ground state $|SG\rangle$ ($\tilde{P}_{{SG}}$) and the state $|G\rangle$ ($\tilde{P}_{{G}}$) reach 0 when the parametric drive is turned off [see the blue-solid
	and green-dotted curves in Fig.~\ref{figS1}(c)].} The entanglement cost (characterized by the
logarithmic negativity $\tilde{E}_{{N}}$) decreases to a low value, i.e., $\tilde{E}_{{N}}\sim 70\%$.
That is, the state of the system after turning off the parametric drive is unpredictable.

\end{widetext}

\end{document}